\documentclass[rmp,twocolumn]{revtex4}
\usepackage{graphicx}
\begin{document}
\title{Recoverable One-dimensional Encoding of Three-dimensional Protein Structures} 

\author{Akira R. Kinjo}
\email{akinjo@genes.nig.ac.jp}
\author{Ken Nishikawa}
\affiliation{Center for Information Biology and DNA Data Bank of Japan,
National Institute of Genetics, Mishima, 411-8540, Japan,
Department of Genetics, The Graduate University for Advanced Studies 
(SOKENDAI), Mishima, 411-8540, Japan}

\begin{abstract}
Protein one-dimensional (1D) structures such as 
secondary structure and contact number provide intuitive pictures
to understand how the native three-dimensional (3D) structure of a protein 
is encoded in the amino acid sequence. However, it has not been clear whether 
a given set of 1D structures contains sufficient information for recovering 
the underlying 3D structure. Here we show that the 3D structure of a protein 
can be recovered from a set of three types of 1D structures, namely, 
secondary structure, contact number and 
residue-wise contact order which is introduced here for the first time.
Using simulated annealing molecular dynamics simulations, the structures 
satisfying the given native 1D structural restraints were sought for 16 
proteins of various structural classes and of sizes ranging from 56 to 146 
residues. By selecting the structures best satisfying the restraints, 
all the proteins showed a coordinate RMS deviation of less 
than 4\AA{} from the native structure, and for most of them, the deviation 
was even less than 2\AA{}. 
The present result opens a new possibility to protein structure prediction
and our understanding of the sequence-structure relationship.
\end{abstract}
\maketitle
\section{Introduction}
Deciphering how the three-dimensional (3D) structure of a protein is 
encoded into the corresponding amino acid sequence is a fundamental step
toward understanding a wide spectrum of complex biological phenomena.
One approach to this problem is to develop a method for structure prediction,
and to interpret the encoding scheme in terms of model parameters and 
optimization algorithms. However, \textit{de novo} or \textit{ab initio} 
methods for 3D structure prediction are often too complicated to clarify the 
relation between sequence and structure. 

One-dimensional (1D) structure prediction \citep{Rost2003} 
is a more intuitive route to understanding the sequence-structure 
relationship.
1D structures are 3D structural features projected onto strings of 
residue-wise structural assignments \citep{Rost2003}, which 
include secondary structures (SS), solvent accessibility and contact 
numbers (CN).
Although 1D structures can show intuitive correspondence between amino acid 
sequence and protein structure, it has not been known whether a given 
set of 1D structures is sufficient for uniquely specifying the underlying 
3D structure.
Clearly, SS alone cannot specify the 3D structure of a globular protein.
Using SS and/or other 1D structures such as CN, is it possible at all 
to recover the native structure? The recent remarkable result by 
\citet{PortoETAL2004} suggests that the answer is affirmative.
They have shown that the principal eigenvector of the contact map of a protein
is essentially equivalent to the contact map itself \citep{PortoETAL2004}.
Using the correct contact map, we can safely recover the native 3D 
structure \citep{VendruscoloETAL1997}.
However, when the principal eigenvector is to be used for reconstructing 
the contact map using the algorithm by \citet{PortoETAL2004}, 
the following strict conditions must be met.
First, the principal eigenvector must be extremely accurate. 
Second, very strict definitions
for residue-residue contact (such as those based on an all-atom 
representation) must be used.
Third, the target protein must be compact and consist of a single domain.
Lack of one of these conditions will result in combinatorial explosion.
It should be also noted that, although the principal eigenvector shows 
a significant correlation with the contact number vector, it is difficult to 
interpret its geometrical meaning. Therefore, it is desirable to find 
1D structures which are more robust, easier to understand, but still 
sufficient for the reconstruction of the native 3D structure.

\citet{KabakciogluETAL2002} have shown that
the number of 3D structures that satisfy the native CN
is limited. The contact number $n_i$ of the $i$-th residue is defined as 
$n_i = \sum_{j}C_{i,j}$ where $C_{i,j}$ is the 
contact map of the native structure of a protein. That is, $C_{i,j} = 1$ if 
the residues $i$ and $j$ are in contact, and $C_{i,j}=0$ otherwise.
In our preliminary study, we constructed many 3D structures that satisfy 
the native SS and CN for a small all-$\alpha$ protein, and found that 
a few percent of the structures were highly 
native-like \citep{KinjoETAL2005}, supporting the result by 
\citet{KabakciogluETAL2002}.
However, we have also found that it is difficult to recover 
the native structures of larger proteins or those with complex topologies 
using only SS and CN restraints. Therefore, either some very powerful 
optimization techniques or other types of 1D structures seemed necessary.

\begin{figure}
\centerline{\includegraphics[width=8cm,clip]{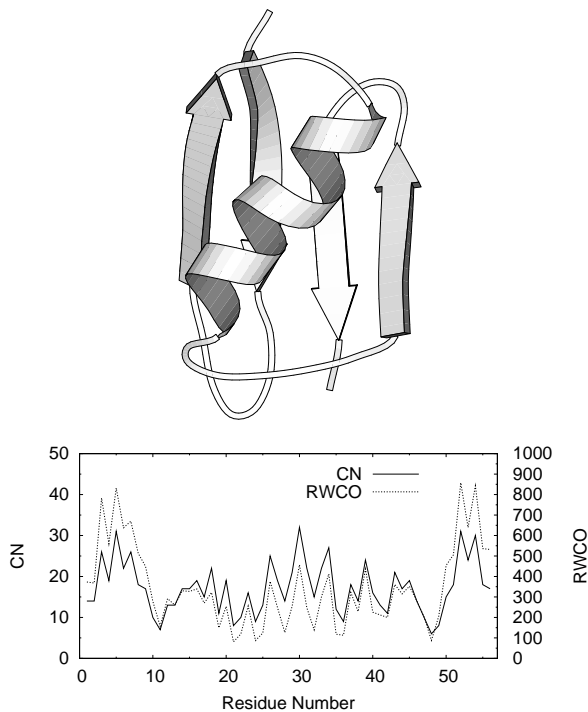}}
\caption{\label{fig:ex}An example of contact number (CN) and residue-wise contact order (RWCO). 
The MolScript \citep{MolScript} drawing in the upper panel shows the native 
fold of Protein G (2gb1), in the bottom panel is the corresponding CN (solid line, left ordinate) 
and RWCO (dashed line, right ordinate).}
\end{figure}

In this paper, we introduce a new kind of 1D structure called residue-wise
contact order (RWCO), and show that, given the native SS, CN and RWCO, 
it is possible to recover the native 3D structures of proteins of 
various topologies. 
The contact order was originally introduced
to quantify the complexity of the native topology of
proteins to investigate the correlation between the 
native structure and its folding rate \citep{PlaxcoETAL1998}. 
As such, the contact order is a per-protein quantity.
Here, we extend the definition of the contact order to make it a per-residue
quantity.
Using the same notation as the definition of CN, the residue-wise contact 
order $o_i$ of the $i$-th residue is defined by $o_i = \sum_{j}|i-j|C_{i,j}$. 
That is, the RWCO of a residue is expressed as the sum of sequence 
separations of contacting residues. An example of CN and RWCO is shown in 
Figure~\ref{fig:ex}. We can see that CN and RWCO exhibit similar trends, 
but the value of RWCO is larger  for the residues making long-range 
contacts (e.g., the N- and C-terminal strands in Figure~\ref{fig:ex}), and 
smaller for those making short-range contacts (e.g., the 
central $\alpha$ helix in Figure~\ref{fig:ex}).
As SS and CN, RWCO has a clear geometrical meaning, and the combination of 
the three types of 1D structures is expected to be more tolerant against small 
perturbations for the reconstruction of 3D structures. 

\section{Materials and Method}
For searching 3D structures that satisfy the given 1D structural 
restraints, we use simulated annealing molecular dynamics simulations. 
In the present paper, two residues are defined to be in contact if the distance
between the $C_{\beta}$ atoms (or $C_{\alpha}$ atoms in case of glycines) is 
less than 12\AA{}. This rather generous cut-off distance has been shown to 
maximize the correlation between predicted and observed contact 
numbers \citep{KinjoETAL2005}. To exclude trivial nearest-neighbor contacts, 
we set $C_{i,j} = 0$ if $|i-j|<3$.
To make CN and RWCO differentiable with respect to atomic coordinates, 
we slightly modify the definition of residue-residue contact 
by using a sigmoid function of inter-residue distance: 
$C_{i,j} = 1/\{1+\exp [w(r_{i,j}-12)]\}$ where $r_{i,j}$ is the distance 
between $C_{\beta}$ atoms of residues $i$ and $j$ \citep{KinjoETAL2005}
(the parameter $w$ determines the sharpness of the sigmoid function, and 
was set to 3 in this paper). 
We used the EMBOSS distance geometry program \citep{EMBOSS}
with default parameters and 
modifications for CN and RWCO restraint functions.
We use an all-atom representation of proteins derived from the AMBER force 
field \citep{AMBER86}. The force field used is the so-called distance 
geometry force field in which all the energy terms are expressed as 
penalty functions including bond lengths, bond angles (1-3 distance),
torsion angles (1-4 distance), short-range (1-4) and long-range (1-5) 
soft repulsions (no attractions) together with chiral center and chiral 
volume restraints \citep{EMBOSS}. Therefore, if a structure perfectly 
satisfies the ideal peptide geometry and all the restraints, the energy value 
should be the minimum value of zero. Disulphide bonds, if any, were ignored, 
and no ligands or co-factors were taken into account.

Secondary structures were assigned by the DSSP program \citep{DSSP}.
For $\alpha$ helices, distance restraints were imposed on hydrogen-bonding 
pairs, and dihedral angle restraints were imposed on $\phi$ and $\psi$ angles.
For $\beta$ strands, distance restraints were imposed between $C_{\alpha}$ 
atoms within each strand segment, and loose dihedral angle restraints for 
$\phi$ and $\psi$ angles were also included. 

Given a set of native contact numbers $\{\hat{n}_{i}\}$, 
the CN restraints were imposed as $w_{n}\sum_{i}(n_{i} - \hat{n}_{i})^{2}$ where
$w_{n}$ is a weight factor which was set to 5. Similarly, with the 
native  residue-wise contact order $\{\hat{o}_{i}\}$,  the RWCO restraints 
was imposed as $w_{o}\sum_{i}(o_{i} - \hat{o}_{i})^{2}$
with the weight factor of 0.5 divided by the sequence length.

To construct a structure, we first generated a random coil which was 
minimized by 500 steps of the conjugate gradient method. Then 
a canonical molecular dynamics simulation at a temperature of 
1000K was performed for 10000 steps, after which the system was cooled by 
2K per 100 steps until the temperature was 100K. Then, the system 
was further cooled by 1K per 100 steps down to 10K. 
The molecular dynamics simulations were performed in four-dimensional space to 
relax the multiple minima problem \citep{Havel1991,EMBOSS}.
Finally, conjugate gradient minimization was applied for 2000 steps to 
recover the structure in three-dimensional space. 
This procedure was iterated for 300 times with different initial random 
coils to yield 300 independent structures for each target protein. 
We sorted these structures 
in increasing order of their total energy to select the best 100 structures.

As target proteins, we chose from the Protein Data Bank \citep{PDB} four all-$\alpha$, four all-$\beta$, 
five $\alpha+\beta$, and three $\alpha/\beta$ proteins whose sequence 
lengths range from 56 to 146 residues (Table~\ref{tab:res}, first column). 
These structures were arbitrarily selected but so as to include proteins 
of varying structural classes and sizes.

\section{Results and Discussion}

\begin{table}
\renewcommand{\baselinestretch}{1.0}
\caption{\label{tab:res}Summary of 3D structures recovered from 1D structures.$^{a}$}
{\begin{tabular}{lrrrrr}\hline
& \multicolumn{3}{c}{\#/RMSD range$^c$} & minimum & minimum \\
Protein (len)$^b$ & $[0,2)$ & $[2,4)$ & $[4,6)$ & energy$^{d}$ & RMSD$^{e}$ \\\hline 
\multicolumn{6}{l}{all $\alpha$}\\
1r69 (63) &  59 &   0 &  37 &   0.6 ( 1.4) &  0.5 ( 1.5) \\
1utg (70) & 100 &   0 &   0 &   0.6 ( 1.7) &  0.5 ( 1.9) \\
256bA (106) &  40 &   0 &   0 &   0.5 ( 2.7) &  0.5 ( 3.0) \\
1mba (146) &   8 &   2 &  15 &   0.5 ( 3.7) &  0.5 ( 3.7) \\
\hline
\multicolumn{6}{l}{all $\beta$}\\
1shg (57) &  19 &   7 &   4 &   0.7 ( 1.5) &  0.7 ( 1.5) \\
1csp (67) &  13 &   4 &   4 &   1.2 ( 2.1) &  1.2 ( 2.1) \\
1ten (89) &   2 &   2 &   2 &   0.9 ( 2.9) &  0.9 ( 2.9) \\
2pcy (99) &   0 &   5 &   2 &  10.9 ( 4.9) &  3.5 ( 9.9) \\
\hline
\multicolumn{6}{l}{$\alpha + \beta$}\\
2gb1 (56) &  21 &   2 &  25 &   0.8 ( 1.3) &  0.7 ( 1.5) \\
1ctf (68) &  88 &   0 &  12 &   0.7 ( 1.4) &  0.7 ( 1.6) \\
1vcc (77) &   3 &  18 &  37 &   2.1 ( 2.3) &  1.6 ( 3.7) \\
2acy (98) &   2 &   2 &   1 &   1.0 ( 2.6) &  0.9 ( 3.4) \\
135l (129) &   0 &  16 &  25 &   3.4 ( 7.3) &  3.1 (11.8) \\
\hline
\multicolumn{6}{l}{$\alpha / \beta$}\\
1ay7B (89) &  20 &   7 &  17 &   0.8 ( 2.3) &  0.6 ( 2.5) \\
1thx (108) &   3 &   5 &   8 &   1.8 ( 4.1) &  0.9 ( 4.2) \\
3chy (128) &   2 &   2 &   1 &   0.5 ( 3.4) &  0.5 ( 3.4) \\
\hline
\end{tabular}}\\
    \begin{flushleft}
$^a$Out of 300 generated structures, 100 lowest energy structures
were selected for the statistics.\\
$^b$PDB identifier with sequence length in parentheses.\\
$^c$ Number of structures resulted in the given range of RMSD (\AA{})
from the native structure. The notation ``$[x,y)$'' indicates the RMSD 
greater than or equal to $x$\AA{} and less than $y$\AA{}.\\
$^d$RMSD (\AA{}) of the structure of the lowest energy with
energy value (no physical unit) in the parentheses.\\
$^e$The minimum RMSD (\AA{}) with energy value (no physical unit) in the 
parentheses.
    \end{flushleft}
\end{table}

For 14 out of the 16 target proteins, we obtained reconstructed 
structures whose $C_{\alpha}$ root mean square deviations (RMSD) from 
the native structure are less than 2\AA{} 
(Table~\ref{tab:res}, second to fourth columns). 
Many of them exhibit even less than 1\AA{} RMSD. For two other targets, 
namely 2pcy (plastocyanin) and 135l (turkey egg white lysozyme),
we still find structures less than 3.5\AA{} RMSD. 
By selecting the structures of the lowest energy,
we can almost always identify highly native-like structures 
(Table~\ref{tab:res}, fifth column). One exception is 2pcy (plastocyanin), 
whose ``best'' structure shows 10.9\AA{} RMSD. However, 
this structure is actually the mirror image of the native structure.
Applying the mirror image transformation to this structure, its 
RMSD from the native structure is 1.4\AA{}. 
Occurrence of mirror image structures is an inherent 
problem of methods which use distance-based restraints 
(CN and RWCO are based on inter-atomic distances). Nevertheless,
the result for 2pcy suggests that it is also possible to obtain structures 
with less than 2\AA{} RMSD if we generate a sufficiently large number of 
structures. 

The minimum RMSDs are shown in the rightmost column of Table~\ref{tab:res}.
These structures do not always correspond to those with 
the lowest energy. Since the average values of the total energy, over 300 
structures generated, are greater by one or two orders of magnitude, 
most of the minimum RMSD structures are significantly close to the 
lowest energy.

The yield of native-like structures greatly varies depending on the target
protein. The native fold of 1utg (uteroglobin) is a very simple one with 
four relatively short $\alpha$ helices, and all the 100 selected 
structures are within 2\AA{} RMSD from the native structure. 
On the contrary, only a handful of native-like structures were obtained for 
2pcy (plastocyanin) which has a complex $\beta$ sandwich topology. 
In general, it seems to be more difficult to obtain native-like
structures for proteins with a large number of long-range contacts.

A reason for the relatively low yield of native-like structure is the use of 
a simple simulated annealing method for the optimization. 
Since all the native-like
structures with less than 2\AA{} RMSD exhibit low energy values, 
the restraints used are 
sufficient for specifying the native-like structures, but many structures are
trapped in local minima during optimization. In fact, we observed 
that setting a high temperature in the initial phase of simulated annealing 
increased the yield of native-like structures. Therefore, the yield is expected
to be even higher if we apply more powerful optimization techniques or 
improved algorithms. 

As can be seen in Figure \ref{fig:ex}, CN and RWCO are highly correlated 
with each other.
Are they both required to reconstruct the native structures?
Performing calculations without using RWCO but following exactly 
the same protocol as above, the total number of native-like structures 
was much smaller (Table \ref{tab:res2}, values before ``/''). 
We obtained native-like structures only for small and/or simple 
proteins such as 1r69, 1utg, 256bA, or 1ctf.
The optimized structures for larger proteins such as 1mba tended to form only 
relatively short-range contacts. Furthermore, even if the correct native
structures were recovered, it was difficult to discriminate them 
by the penalty function. A slightly better, but qualitatively similar result 
was obtained when CN was omitted in the calculations 
(Table \ref{tab:res2}, values after ``/''). In this case, compared to the case 
without RWCO, the optimized structures tended to contain a comparable or 
smaller number of contacts, but of longer range.
From these observations, we conclude that CN and RWCO contain complementary 
information required to accurately determine the native-like structures.

\begin{table}
\caption{\label{tab:res2}Summary of 3D structures recovered from 1D structures without RWCO (values before ``/'') or without CN (values after ``/'') (cf. Table \ref{tab:res}).}
{\begin{tabular}{lrrrrr}\hline
& \multicolumn{3}{c}{\#/RMSD range} & minimum & minimum\\
Protein & $[0,2)$ & $[2,4)$ & $[4,6)$ & energy[\AA{}]  & RMSD[\AA{}]\\\hline 
1r69  &  6 / 15 & 15 / 11 &  4 / 15  &  1.3 /  1.2  &  1.2 /  0.8 \\
1utg  &  2 / 23 & 31 / 56 & 10 /  3  &  2.0 /  0.9  &  1.7 /  0.8 \\
256bA &  1 / 14 &  8 /  3 &  2 /  0  &  8.8 /  2.1  &  1.6 /  1.3 \\
1mba  &  0 /  0 &  0 /  4 &  0 /  3  & 13.3 /  2.3  & 10.4 /  2.3 \\
\hline
1shg  &  0 /  0 &  0 /  2 &  1 /  6  &  8.6 /  9.7  &  4.1 /  2.7 \\
1csp  &  0 /  0 &  1 /  2 &  4 /  2  & 10.0 /  9.9  &  2.8 /  2.9 \\
1ten  &  0 /  0 &  0 /  0 &  1 /  0  & 10.4 / 13.3  &  5.9 /  8.0 \\
2pcy  &  0 /  0 &  0 /  0 &  0 /  0  & 13.3 / 13.2  &  8.2 /  7.6 \\
\hline
2gb1  &  0 /  0 &  0 /  0 &  2 /  1  &  6.9 /  7.5  &  5.1 /  5.9 \\
1ctf  & 11 / 21 &  2 /  6 &  7 /  6  &  1.5 /  1.1  &  1.2 /  0.9 \\
1vcc  &  0 /  0 &  0 /  0 &  3 /  1  & 10.8 / 12.0  &  5.0 /  5.3 \\
2acy  &  0 /  0 &  0 /  0 &  1 /  1  & 12.4 / 13.2  &  5.7 /  5.4 \\
135l  &  0 /  0 &  0 /  0 &  0 /  0  & 13.3 / 14.8  & 10.5 /  8.5 \\
\hline
1ay7B &  0 /  0 &  0 /  0 &  0 /  1  & 10.2 / 10.2  &  6.2 /  5.4 \\
1thx  &  0 /  0 &  0 /  0 &  0 /  0  & 12.4 /  9.1  &  7.4 /  7.1 \\
3chy  &  0 /  0 &  0 /  0 &  0 /  0  & 14.9 / 12.0  &  6.6 /  9.9 \\
\hline
\end{tabular}}{%
}
\end{table}

It is of interest to ask 
whether SS, CN and RWCO uniquely specify the native 3D structure of a protein 
(except for the mirror image). We expect such is the case, although we cannot 
give the definite conclusion based on the restraint-based, rather than 
constraint-based, method as used in this study. All the optimized structures 
do satisfy the given 1D structural restraints to a certain extent, but those 
with high energies tend to contain significant distortions in 
their local geometry and large steric overlaps. Thus, given the native SS, CN
and RWCO, the number of the structures consistent with these restraints as 
well as the ideal peptide chain geometry should be very limited. 
It should be noted that this argument probably applies only if the full-atom 
representation is used, otherwise there may exist non-native-like structures 
with low energy values.

Although we have performed a direct optimization of 3D structures by  
imposing 1D structural restraints, it may be also possible to 
first reconstruct the contact map satisfying the 1D restraints, and then 
recover the 3D structure from the contact map. In an initial phase of the 
present study, we applied a deterministic depth-first search algorithm similar
to that of \citet{PortoETAL2004}. However, this method failed to 
converge. Since both CN and RWCO are accumulative quantities, there may not be 
any strategy to efficiently eliminate unsuccessful candidates in early stages 
of the search. Another possibility is applying a Monte Carlo method in 
contact map space. We have applied a variant of the multicanonical 
methods \citep{WangANDLandau2001}, but failed to 
find a solution exactly satisfying the 1D restraints. Nevertheless, 
for small proteins, thus obtained contact maps that best, but not exactly, 
satisfy the restraints contained at least 30 to 40\% of the correct native
contacts, and appeared similar to the native contact map by visual inspection.
Therefore, it may be possible to use such contact maps to construct starting 
conformations for further optimizations.

Since the three types of 1D structures, SS, CN and RWCO, are sufficient 
for determining the native 3D structure, it is possible to predict the 
native structure of a protein if we can accurately predict 
these 1D structures. Methods for secondary structure prediction are now 
quite mature and are already routinely used in \textit{de novo} 3D structure 
prediction \citep{Rost2003}. We have previously developed a method to predict 
CN from amino acid sequence to a decent accuracy with
a correlation coefficient of 0.63 \citep{KinjoETAL2005}. 
We have recently developed a simple linear regression method for 
RWCO prediction which yields a moderate correlation of 0.59 between 
the predicted and native RWCOs \citep{KinjoANDNishikawa2005b}.
At present, we do not expect that the native 3D structure can be 
obtained by using the predicted 1D structures: 1D predictions of 
higher accuracies must be achieved.
Nevertheless, if the accuracies of 1D structure prediction are
sufficiently improved, the missing link 
between amino acid sequence and the native 3D structure of globular
proteins may be completed.

\begin{acknowledgments}
We thank Takehiro Nagasima for valuable comments. 
Most of the computations were carried out at the supercomputing facility of 
National Institute of Genetics, Japan. This work was supported in part by a 
grant-in-aid from the MEXT, Japan.
\end{acknowledgments}


\end{document}